# Selective linking from social platforms to university websites: a case study of the Spanish academic system

**Enrique Orduña-Malea[1*] and José-Antonio Ontalba-Ruipérez[1]**

[1]Department of Audiovisual Communication, Documentation, and History of Art. Polytechnic University of Valencia (UPV), Valencia, Spain.
Camino de Vera s/n, Valencia 46022, Spain.
* e-mail: enorma@upv.es

**Abstract** Mention indicators have frequently been used in Webometric studies because they provide a powerful tool for determining the degree of visibility and impact of web resources. Among mention indicators, hypertextual links were a central part of many studies until Yahoo! discontinued the 'linkdomain' command in 2011.
Selective links constitute a variant of external links where both the source and target of the link can be selected. This paper intends to study the influence of social platforms (measured through the number of selective external links) on academic environments, in order to ascertain both the percentage that they constitute and whether some of them can be used as substitutes of total external links.
For this purpose, 141 URLs belonging to 76 Spanish universities were compiled in 2010 (before Yahoo! stopped their link services), and the number of links from 13 selected social platforms to these universities were calculated. Results confirm a good correlation between total external links and links that come from social platforms, with the exception of some applications (such as Digg and Technorati). For those universities with a higher number of total external links, the high correlation is only maintained on Delicious and Wikipedia, which can be utilized as substitutes of total external links in the context analyzed. Notwithstanding, the global percentage of links from social platforms constitute only a small fraction of total links, although a positive trend is detected, especially in services such as Twitter, Youtube, and Facebook.

## 1. Introduction

As a metric and quantitative discipline, Webometrics has traditionally paid particular attention to the identification, definition, and application of web-based indicators, and their possible correlation with other kinds of non-web indicators.

In this regard, categories of web indicators have been proposed by several authors, such as Dhyani et al. (2002; cited by Faba et al. 2004), Alonso et al. (2003), or Aguillo (2009a; 2009b). However, the most important works related to web indicators were the result of various projects financed with European funds, specifically the project WISER (Web Indicators for Science, Technology & Innovation Research[1], with emphasis on the corresponding website, Indicators Web Portal[2]), and the project EICSTES (European Indicators, Cyberspace and the Science-Technology-Economy System)[3].

Among the indicators studied in the aforementioned projects, the measurement of mentions (also known as invocations) are of particular relevance, due to their ability to measure impact, popularity, and usefulness of digital and web resources (Cronin et al. 1998).

The main goal of this study is to analyze the scope of one type of web mention (hypertextual selective linking), to illustrate the level of influence of social networks and resource-sharing systems in an academic environment.

For this purpose, the following section will propose a classification of mention indicators. The section after that will focus on hypertextual mentions, highlighting the value and importance of these link mentions, proposing a classification of them according to the source link, and noting the existence of selective linking (whereby the placement of the source link can be chosen). Finally, the section on selective linking will focus on the hypertextual links that come from social platforms, and some resource sharing applications.

## 1.1. Mention indicators

The following classification of mention indicators is proposed:

**a) Text mentions (invocations)**

This set of indicators relates to the quantification of the number of times that a particular string of characters is displayed on files hosted in the explored network space.

These strings can represent certain concepts, phrases, authors, publications, citations, complete documents, or even certain web sites and online seats. Therefore, they are indicators that characterize the impact of the content published on the Web, based on the number of times they are referenced by other documents on the Net.

**b) Query mentions (search terms)**

These measurements refer to the set of strings that have led a user to visit a particular site or seat online. Therefore, they try to characterize the query set (and the number thereof) through which an online site is visited.

**c) Description mentions (metadata)**

Strings used to describe the content published on the Web. All kinds of metadata, keywords, descriptors and labels used to describe online content belong to this category.

**d) Hypertext mentions (hyperlinks)**

In this paper, hyperlinks are registered as a typology of mention indicators, in which the string of characters (both the destination URL and the text that serves as a link) is made explicit by a specific markup language.

Since hyperlinks (or just links) are the indicators that most lines of research (both descriptive and applied) have generated, a further analysis of them is necessary.

## 1.2. The value of hypertext mentions

Link analysis has opened up a wide variety of research areas, most notably the following:

*a) Types of link*

In general terms, the description of a link takes into account the characteristics of the source, the target, and the purpose of the link.

With regard to the link generator document (source), this description falls under the following categories (Garcia-Santiago 2001):

- Intrinsic link: refers to other places on the same webpage, and are identified by various HTML tags.

- Internal link: connects, via a URL, to other pages or files placed on the same site, either to complete the same document (for example, an image), or to refer to another separate but related website document, facilitating browsability.

- External link: refers to pages or files located on another website.

With regard to the link receiver document (target), the categories are (Björneborn and Ingwersen 2004):

- Incoming link (inlink): can be both internal and external.
- Outgoing link (outlink): can be both internal and external.
- Self-link: the source (source) and destination (target) match.

Finally, with regard to its purpose, there are basically two types of link in hypertext documents (Baron et al. 1996):

- Organizational links: links that organize information in a website: indicate the physical location of hypertext documents (for example: "Next Page").

- Content-based links: related to the significance of hypertext information (e.g. "More Information").

Deeper analysis of the different purposes of linking behavior leads to the following area of research, focused specifically on assessing the value of hyperlinking:

*b) The meaning of links*

It is logical to assume that a document (or other unit of study) containing quality information will be linked to (mentioned) by more pages than another document with less valuable information. For this reason, quantifying the number of links (mostly external) that a document receives is a sign of its importance, or at least its impact on the environment (Heylighen 2000).

However, the reasons for which an author or community (Gibson et. al 1998a; 1998b) generate a hyperlink in a document to another document are not easy to establish.

Kim (2000) investigated the motivations of authors for creating links to electronic journal articles, finding that many links were made simply to allow access to electronic resources. Smith (1999), Thelwall (2001), and Bar-Illan (2003; 2005) have also analyzed links in academic environments, determining the existence of links motivated by other reasons (such as referring to educational or informative materials), which reduces the accuracy of link count to studying scientific relations.

In academic environments, links are sometimes created for relatively trivial reasons, for example, an academic page may contain links related to the author's hobbies or family and friends. In addition, many links between pages on a website are created primarily for the purpose of navigation within the site (aforementioned as organizational links).

Among the abundant literature that has previously addressed this issue, special attention should be paid to the preliminary works done by Hass and Grams (1998), who proposed a pioneering link taxonomy, and later, by Wilkinson et al (2003), centered on the UK university system. Particular consideration should also be given to the classic works of Bar-Illan (2003; 2005), who proposed a detailed classification of hyperlinks by disaggregating the sources, targets, and links into several categories and subcategories, and applying this classification scheme to a study of the Israeli university system.

More recently, Thelwall (2009) exhaustively analyzes the nature of the links and the reasons behind the creation of links in different areas, and concludes that the links are "not a perfect source of evidence because not all hyperlinks have been carefully created by the authors, considering what page should be most appropriate for the target of the link". Seeber et al. (2012) also analyze factors pertinent to web links within European Higher Education Institutions, providing a further detailed description of the literature related to the different motivations for linking in academic environments.

*c) The analysis of links*

Thelwall (2010) proposes two main areas of study within link analysis: impact assessment, and mapping relationships.

- Assessing the impact of the links: this line of research focuses on the compilation of a set of web sites and subsites (especially if they are also online sites), and the subsequent comparison between them according to the number of external links to each one of them.
- Building relationship maps through links: these maps include organizational relationships based on links, and maps of subject areas or organizations, whose aim is to establish some kind of similarity of content.

The limitations of link analysis (due to the problems of understanding their motivations) have led several authors (Katz 2004) to more complex analysis to obtain more accurate results from which to draw more precise conclusions about relationships between websites.

In this sense, co-links and webometric coupling have been widely and successfully used in specific environments. Larson (1996) performed one of the first co-link studies applying multivariate techniques to analyze the intellectual structure of the Web beyond simple hierarchical directories (at the time), such as Yahoo!.

### 1.3. A general classification of linkage indicators

A classification for hypertext mentions (as discussed in section 1.1d) is presented below, with due consideration given to external and internal links:

*a) General linking*

The links are counted regardless of the source and target. They can be inlinks or outlinks with respect to the unit of study under consideration, and internal or external whether source and destination match or not.

*b) Weighted linking*

The posted links are weighted by the importance given to each of them by page ranking sites, such as PageRank, Domain mozRank or Compete Rank.

*c) Selective linking*

Links are counted if they are directed to or from a particular online site. This may be a TLD (domain linking) or an online site or subsite (site linking). The latter can be divided into products (e.g., a repository or a platform) or institutions (e.g., a university).

### 1.4. Selective linking from social platforms

Selective links provided from social platforms represent an important aspect, due to increasing user activity in these services. The amount of data within these platforms is now starting to be used in order to quantify and evaluate scientific activity (Cabezas-Clavijo and Torres-Salinas 2012).

In this sense, initiatives such as Altmetrics[6] try to design and implement alternative metrics that "based on a diverse set of social sources could yield broader, richer, and timelier assessments of current and potential scholarly impact".

Among the new services which are starting to be used as a new basis for research into impact evidence, the following are particularly relevant (Priem and Hemminger 2010): bookmarking (such as Delicious), reference managers (such as Mendeley), recommendation systems (such as Digg), blogging (technorati), microblogging (Twitter), and social networks (Facebook).

These platforms and associated new indicators are putting scientometrics one step forward, as quantitative works with Wikipedia (Nielsen 2007), Twitter (Priem and Costello 2010; Thelwall and Priem 2012), and Mendeley (Li et al. 2011), among others, show.

In any case, a void in the literature has been detected precisely in the quantification of links from these platforms to academic spaces (i.e., selective linking), such as universities, an important issue if we consider the power of links to analyze universities' websites, as shown in previous sections.

## 2. Objectives

The main objective of this study is to analyze the influence both of social platforms and resource sharing applications in the web visibility of the academic environment -in particular the Spanish University system- in 2010. For this reason, the specific objectives are as follows:

- To calculate the number of links that a selected group of social platforms provide to Spanish universities.
- To analyze the correlation between these selective links and the total external links received in the same period by these universities.
- To find out if some selective links from any social platform considered can be used as a good substitute of total external links, and therefore, can accurately reflect the web visibility of the universities.

## 3. Method

### 3.1. Data gathering

The analysis is applied to the Spanish University System, consisting, in 2010, of 76 universities, both public and private. Both the list of universities and their URLs were obtained during the last quarter of 2009, through the following official sources: Ministry of Education[4], and Conference of Rectors of Spanish Universities[5].

Once the different universities were compiled and classified, the URL for each one of them was identified as well. In addition to the official URLs (those indicated in the official sources consulted), it was found that alias and alternative domains existed at various universities. To conduct the study it was determined that all of these URLs be exhaustively compiled, even if they redirected to another URL, or just did not have enough information stored.

Alias domains (URLs that share the same second level domain, but have different TLDs) were identified by manually checking, at each university, the existence of the following domains: CAT, COM, EDU, ES, NET, and ORG. Alternative domains (different second level domain) were searched through Yahoo! Site Explorer (due to its usefulness in finding domains). Additionally, the institutional website was explored manually in order to find possible alternative URLs.

### 3.2. Data measurement: sources and indicators

All sources utilized for selective and external linking are described below (all of them accessed on 01/05/2011):

*Delicious*
*http://delicious.com*

From Yahoo! (owner of Delicious) the query "linkdomain:domain.tld site:delicious.com" cannot directly be made. For that reason, the Delicious internal search engine is employed directly by using the command "**site:**domain.tld", which returns the total number of bookmarks which link to each academic URL (figure 1).

Figure 1. Example of query on Delicious (2010)

Today this query is no longer supported, and the number of bookmarks pointing to a specific URL from Delicious cannot be quantified.

*Yahoo! search*
*http://search.yahoo.com*

The advanced settings of the browser are used to perform the following preliminary operations: deactivation of the parental control filter, and setting 100 items as the maximum number of results per page.

During the measurement period (2010), due to the merger between Yahoo! and Bing, the U.S. database did not provide data links or some combined queries. For that reason the Spanish database was used. However, as of April 2011, these commands also stopped working properly in the Spanish version and in November 2011 Yahoo! discontinued the service entirely, so that such queries are no longer possible today.

All indicators (selective linking; external linking), commands (linkdomain; site) and sources (Delicious; Yahoo! Search) are summarized in table 1. The page count indicator is also provided with the purpose of contextualizing results according to the size of web domains.

The selective sources of the selected links are the following: Academia; Delicious; Digg; Facebook, Flicker; Linkedin; Meneame; Sinc platform; Slideshare; Technorati; Twitter; Wikipedia, and Youtube. Source selection is exploratory and not exhaustive, but includes the most popular social services currently in use. The use of Meneame and Sinc platform are justified by their importance for the Spanish user base. Table 1 also shows the corresponding URLs for each service.

Table 1. Indicators, commands and sources

| INDICATOR | SOURCE | COMMAND |
|---|---|---|
| SITE LINKING | Yahoo! Search | **linkdomain:**domain.tld **site:**academia.edu |
| | | **linkdomain:**domain.tld **site:**digg.com |
| | | **linkdomain:**domain.tld **site:**facebook.com |
| | | **linkdomain:**domain.tld **site:**flickr.com |
| | | **linkdomain:**domain.tld **site:**linkedin.com |
| | | **linkdomain:**domain.tld **site:**meneame.net |
| | | **linkdomain:**domain.tld **site:**plataformasinc.es |
| | | **linkdomain:**domain.tld **site:**slideshare.net |
| | | **linkdomain:**domain.tld **site:**technorati.com |
| | | **linkdomain:**domain.tld **site:**twitter.com |
| | | **linkdomain:**domain.tld **site:**wikipedia.org |
| | | **linkdomain:**domain.tld **site:**youtube.com |
| | Delicious | **site:**domain.tld |
| EXTERNAL LINKING | Yahoo Search | **linkdomain:**domain.tld –**site:**domain.tld |
| COUNT PAGE | Yahoo Search | **site:**domain.tld |

Finally, it should be clarified that, although the nomenclature used previously for the online representation of each university is their institutional URL, the selected indicators are applied not only for that single URL but for all sub-folds, sub-domains and files under the official registered web domain. For that reason, when a URL is mentioned, the set of URLs under that web domain should be interpreted.

### 3.3. Data measurement: samples

All measurements were taken quarterly in 2010, in the following months: March, June, September, and December. In order to avoid differences due to geographical reasons (accessing different databases), all measures were done through the same IP: 81.202.220.158.

### 3.4. Data measurement: statistical analysis

Raw data of all URLs are normalized from 0 to 100 by a transformation process (Rocki 2005), with the aim of working subsequently with the mean relative representation factor in visibility (Rv).

To this end, the sum of all links obtained from all the URLs of the whole Spanish system over an entire month (accumulated visibility) is considered equal to 100, and the value of each URL is calculated proportionately:

$$v_{vn} = \frac{x_{vn}}{\sum_{i=1}^{N} x_{vn}} \cdot 100 \; ; \; \text{[equation 1]}$$

$v_{vn}$ = Normalized value obtained in visibility (v) for a URL (*n*).
$x_{vn}$ = Raw valued obtained in visibility (v) for a URL (*n*).
N = Set of URLs considered.

After normalizing results, a percentage proportional to the total visibility (total inlinks received from a particular social site to all Spanish universities) is obtained for all universities for every monthly measurement (4 snapshots in this case), a concept called "relative representation". These values are displayed in tables 3-15.

Then, the average of $v_{vn}$ is calculated monthly, obtaining a value, also between 0 and 100, which is referred to as the "mean relative representation factor in visibility (Rv)". This factor can be calculated for any set of selected sites and any period of time (Orduna-Malea et al. 2010).

$$R_v = \frac{\sum_{i=1}^{M} v_{vn}}{M} \; ; \; \text{[equation 2]}$$

Where M is the number of months analyzed (in this case 4 samples, for March, June, September and December 2010).

Finally, in order to calculate the growth rate of web domains during the period, r (%), the compound interest formula was used:

$$A = P\left(1 + \frac{r}{n}\right)^{n \cdot t} \; ; \; \text{[equation 3]}$$

A = Accumulated count after n years.
P = Principal amount.
r (%) = Annual rate of growth.
n = Number of times the growth is compounded per year.
T = Number of years.

All data gathered were exported to a spreadsheet to be statistically analyzed. Additionally, the XLStat application was used to obtain Spearman correlations between indicators.

## 4. Results

All results gathered are separated into a descriptive analysis of each platform, and a correlation analysis of the corresponding results.

### 4.1. Descriptive analysis

Descriptive results are aggregated by site (platform), and university (target domain):

#### 4.1.1. Results per site

From 76 universities, a total of 141 URLs were retrieved, which form the final data sample (all universities and their corresponding URLs are available in Annex 1). Table 2 displays the accumulated links (summation of links from all 141 URLs) for every platform considered, in the four snapshots of gathered data (March, June, September, and December). The total external inlinks are provided as well.

**Table 2. Evolution of accumulated links**

| PLATFORMS | MAR | JUN | SEP | DEC |
|---|---|---|---|---|
| Delicious | 96,299 | 104,549 | 104,596 | **116,985** |
| Wikipedia | 37,402 | 49,543 | 46,574 | **46,508** |
| Facebook | 2,777 | 5,037 | 4,838 | **4,253** |
| Linkedin | 2,089 | 2,385 | 2,690 | **2,946** |
| Sinc | 2,165 | 1,013 | 1,921 | **1,939** |
| Meneame | 5,411 | 1,884 | 1,565 | **1,726** |
| Youtube | 868 | 756 | 1,043 | **1,303** |
| Academia | 1,001 | 1,848 | 1,411 | **1,151** |
| Twitter | 311 | 429 | 518 | **637** |
| Slideshare | 341 | 459 | 555 | **576** |
| Flickr | 330 | 379 | 502 | **557** |
| Digg | 282 | 207 | 222 | **53** |
| Technorati | 85 | 25 | 29 | **36** |
| External inlinks | 8,996,501 | 10,255,083 | 10,057,155 | 10,411,729 |

The external links targeted at the Spanish university system grew from 8,996,501 links in March to 10,411,729 in December, showing an upward trend, which only decreased in September, probably due to the summer season.

The cumulative total number of inlinks in December, counting all the social sites examined, amounts to 178,670 (of which 116,985 correspond to Delicious, and 46,508 are for Wikipedia, constituting the platforms that provide the greatest number of links to Spanish universities). This amount represents only 1.72% of total external links (10,411,729) accumulated by Yahoo! in December.

On the other hand, both Digg and Technorati represent the platforms with the lowest presence of Spanish universities, with only 56 and 36 links respectively. The very popular sites Facebook and Twitter show unexpectedly low results, although they present an increasing trend. Figure 2 shows the evolution over time for each platform

**Figure 2. Accumulated *Social site inlinks* over time (*Yahoo!*)**

### 4.1.2. Results per university

The results confirm that "ucm.es" is the academic domain with the highest number of total external inlinks (485,505 for December), followed by "ugr.es" (375,573), "uv.es" (358,758), "ua.es" (323,115), "usal.es" (316,942), and "uab.es" (310, 856). Annex 2 (table 17) contains the raw data (total inlinks) received per URL from each of the platforms considered, showing only those with at least 100,000 total external inlinks (38 of 141 URLs).

The results per platform also confirm the predominance of Delicious and Wikipedia, where "ucm.es" also constitutes the domain which receives the most links from these two sources. Despite this, some domains perform extremely well on specific platforms. For example, from Delicious, "ehu.es" (3,466), "uam.es" (2,456), and "uc3m.es" (2,833) receive a high percentage of links if compared with total external links. From Wikipedia, "unirioja.es" receives 6,140 links (the second in this ranking after "ucm.es"). This confirms that some universities are expressly focusing their activity on some social platforms, and that activity is not extended to the remaining services.

Annex 3 (table 18) shows average Rv values attained by each URL in each of the platforms (considering only those URLs with an Rv>1). Data is sorted by Rv values taking total external inlinks into consideration. If table 18 is compared with table 17, the same URLs (except "uoc.edu") can be observed in the top ten, with slight changes in some positions (for example "ugr.es" and "uv.es" interchange second and third position). Another case of interesting behavior is detected with "uji.es", which ranks 7[th] according to Rv, but is ranked 27[th] regarding total external links in December due to a significant drop in total external links after June.

These Rv values allow the representativeness of each domain to be determined, avoiding raw data, in the period analyzed. For example, "us.es" achieves high scores in Flickr (12.10), Technorati (11.61), and Wikipedia (13.27), but little representativeness in the remaining platforms. Another similar case is "unirioja.es", which stands out on Academia (10.04) and Wikipedia (11.97).

Other universities perform exceptionally well on only one specific platform; for example, "uc3m.es" and "uam.es" stand out on Sinc Platform; "ie.edu" and "upv.es" on Technorati; and "uib.es" on Digg.

Another interesting case study is that of the "ua.es" and "uji.es" websites, which do not stand out on any of the studied platforms, but get high average Rv in external links (fourth and seventh position respectively). This effect indicates that social platforms constitute a small proportion of global links.

Finally, the effects of URL dispersion among different URLs belonging to the same university should be noted, especially in the Catalan university system. For example: "uab.es" and "uab.cat", "ub.es" and "ub.edu", or "upc.edu" and "upc.es". All of them in the top URLs, but the links are divided into each web domain.

To further illustrate these performance differences, in relation to the origin of the links, a detailed analysis of each platform is provided below.

**a) Social networks**

The main social networking sites are described in this section: Academia (academic network), Facebook (general purpose network), and Linkedin (professional network).

*Academia.edu*

The values are generally quite low (table 3). Only 2 domains ("unirioja.es" and "uv.es") manage over a hundred links in the last snapshot. Also of note is the high position attained by "deusto.es", and the Catalan universities ("uab.es", "upf.edu" and "ub.es" are among the 10 domains with the most links).

Table 3. Inlinks and Rv for Academia.edu

| DOMAINS | LINKS | | | | RELATIVE REPRESENTATION | | | | Rv | r (%) |
|---|---|---|---|---|---|---|---|---|---|---|
| | MAR | JUN | SEP | DEC | MAR | JUN | SEP | DEC | | |
| unirioja.es | 106 | 144 | 169 | 113 | 10.59 | 7.79 | 11.98 | 9.82 | **10.04** | 0.06 |
| uv.es | 128 | 146 | 107 | 116 | 12.79 | 7.90 | 7.58 | 10.08 | **9.59** | -0.10 |
| uab.es | 157 | 87 | 88 | 67 | 15.68 | 4.71 | 6.24 | 5.82 | **8.11** | -0.77 |
| unizar.es | 28 | 125 | 114 | 78 | 2.80 | 6.76 | 8.08 | 6.78 | **6.10** | 1.17 |
| upf.edu | 22 | **283** | 23 | 28 | 2.20 | 15.31 | 1.63 | 2.43 | **5.39** | 0.25 |
| uc3m.es | 49 | 50 | 111 | 66 | 4.90 | 2.71 | 7.87 | 5.73 | **5.30** | 0.31 |
| ucm.es | 46 | 80 | 56 | 63 | 4.60 | 4.33 | 3.97 | 5.47 | **4.59** | 0.33 |
| deusto.es | 12 | 189 | 30 | 30 | 1.20 | 10.23 | 2.13 | 2.61 | **4.04** | 1.03 |
| uvigo.es | 18 | 143 | 36 | 19 | 1.80 | 7.74 | 2.55 | 1.65 | **3.43** | 0.05 |
| ub.edu | 33 | 43 | 52 | 47 | 3.30 | 2.33 | 3.69 | 4.08 | **3.35** | 0.37 |

An inconsistency in the values of "upf.edu" in June should be noted, causing an artificial increase of Rv. This effect leads to some undesirable results. For example, "uv.es" links increase from March to June (from 128 to 146), despite the Rv decrease in June due to the high increase of "upf.edu". This proves that the normalization method used is not valid for evolution purposes (raw data should be used). In any case, the average value of Rv mitigates this effect, and should be used instead of monthly values.

*Facebook*

Table 4 shows the corresponding values for Facebook. Although the number of links is higher than in Academia.edu, they still remain low. In any case, there was a significant increase from March to June, but after then the data decreased slightly.

The domains "ub.edu" and "ucm.es" are those which achieve the highest representative value (more than 500 links on the last snapshot of data). A possible reason could be that they are both larger and well-established institutions, with high enrollment rates. Another point to mention is the good performance of private universities, with "ie.edu", "unav.es", "uoc.edu" and "uem.es" among the top places.

Table 4. Inlinks and Rv for Facebook

| DOMAIN | LINKS | | | | RELATIVE REPRESENTATION | | | | Rv | r (%) |
|---|---|---|---|---|---|---|---|---|---|---|
| | MAR | JUN | SEP | DEC | MAR | JUN | SEP | DEC | | |
| ub.edu | 312 | 700 | 610 | 546 | 11.24 | 13.90 | 12.61 | 12.84 | **12.64** | 0.60 |
| ucm.es | 239 | 505 | 464 | 505 | 8.61 | 10.03 | 9.59 | 11.87 | **10.02** | 0.82 |
| uem.es | 215 | 236 | 223 | 194 | 7.74 | 4.69 | 4.61 | 4.56 | **5.40** | -0.10 |
| uoc.edu | 157 | 246 | 228 | 195 | 5.65 | 4.88 | 4.71 | 4.58 | **4.96** | 0.22 |
| unav.es | 168 | 206 | 199 | 177 | 6.05 | 4.09 | 4.11 | 4.16 | **4.60** | 0.05 |
| ie.edu | 148 | 178 | 191 | 167 | 5.33 | 3.53 | 3.95 | 3.93 | **4.18** | 0.12 |
| usal.es | 124 | 173 | 188 | 152 | 4.47 | 3.43 | 3.89 | 3.57 | **3.84** | 0.21 |
| upc.edu | 62 | 192 | 205 | 173 | 2.23 | 3.81 | 4.24 | 4.07 | **3.59** | 1.17 |
| us.es | 81 | 189 | 166 | 146 | 2.92 | 3.75 | 3.43 | 3.43 | **3.38** | 0.63 |
| unirioja.es | 77 | 179 | 140 | 129 | 2.77 | 3.55 | 2.89 | 3.03 | **3.06** | 0.55 |

*Linkedin*

Table 5 shows links for Linkedin. Although the raw values obtained are very low, a growing trend is detected in the period: from a cumulative 2,089 links in June to 2,946 in December.

Table 5. Inlinks and Rv for Linkedin

| DOMAIN | LINKS | | | | RELATIVE REPRESENTATION | | | | Rv | r (%) |
|---|---|---|---|---|---|---|---|---|---|---|
| | MAR | JUN | SEP | DEC | MAR | JUN | SEP | DEC | | |
| *ie.edu* | 160 | 179 | 185 | 202 | 7.66 | 7.51 | 6.88 | 6.86 | **7.22** | 0.24 |
| *upc.edu* | 143 | 163 | 185 | 195 | 6.85 | 6.83 | 6.88 | 6.62 | **6.79** | 0.32 |
| *upm.es* | 111 | 123 | 139 | 150 | 5.31 | 5.16 | 5.17 | 5.09 | **5.18** | 0.31 |
| *uc3m.es* | 95 | 103 | 112 | 129 | 4.55 | 4.32 | 4.16 | 4.38 | **4.35** | 0.32 |
| *upv.es* | 83 | 99 | 111 | 116 | 3.97 | 4.15 | 4.13 | 3.94 | **4.05** | 0.35 |
| *upf.edu* | 61 | 68 | 73 | 85 | 2.92 | 2.85 | 2.71 | 2.89 | **2.84** | 0.35 |
| *ub.edu* | 57 | 68 | 77 | 82 | 2.73 | 2.85 | 2.86 | 2.78 | **2.81** | 0.38 |
| *unav.es* | 57 | 63 | 76 | 78 | 2.73 | 2.64 | 2.83 | 2.65 | **2.71** | 0.33 |
| *uam.es* | 53 | 63 | 71 | 72 | 2.54 | 2.64 | 2.64 | 2.44 | **2.57** | 0.32 |
| *ugr.es* | 51 | 59 | 71 | 77 | 2.44 | 2.47 | 2.64 | 2.61 | **2.54** | 0.43 |

IE University surpasses 200 links in December and has a corresponding average Rv of 7.22. Also worth pointing out is the presence of polytechnic universities at the top (UPC, UPM and UPV) and good performances in some private universities, such as UNAV, UOC, and slightly behind, the URL.

**b) Social news managers**

The social news services Digg (international) and Meneame (Spain and Latin America) are analyzed below:

*Digg*

The data obtained for Digg (Table 6) are practically nonexistent. The domains "uib.es" and "upf.edu" should be noted but they gradually lost links in each data sample. Also of note are the results of "upc.edu" in September, which returns to its initial state in December.

**Table 6. Inlinks and Rv for Digg**

| DOMAIN | LINKS | | | | RELATIVE REPRESENTATION | | | | Rv | r (%) |
|---|---|---|---|---|---|---|---|---|---|---|
| | MAR | JUN | SEP | DEC | MAR | JUN | SEP | DEC | | |
| *uib.es* | 107 | 82 | 72 | 16 | 37.94 | 39.61 | 32.43 | 30.19 | **35.04** | -1.51 |
| *upf.edu* | 40 | 29 | 26 | 8 | 14.18 | 14.01 | 11.71 | 15.09 | **13.75** | -1.33 |
| *upc.edu* | 8 | 6 | **60** | 7 | 2.84 | 2.90 | 27.03 | 13.21 | **11.49** | -0.13 |
| *ugr.es* | 27 | 17 | 13 | 4 | 9.57 | 8.21 | 5.86 | 7.55 | **7.80** | -1.52 |
| *upm.es* | 16 | 12 | 6 | 1 | 5.67 | 5.80 | 2.70 | 1.89 | **4.02** | -2.00 |
| *uab.es* | 14 | 8 | 5 | 2 | 4.96 | 3.86 | 2.25 | 3.77 | **3.71** | -1.54 |
| *upf.es* | 9 | 9 | 7 | 2 | 3.19 | 4.35 | 3.15 | 3.77 | **3.62** | -1.25 |
| *uma.es* | 4 | 3 | 4 | 2 | 1.42 | 1.45 | 1.80 | 3.77 | **2.11** | -0.64 |
| *udc.es* | 3 | 3 | 2 | 2 | 1.06 | 1.45 | 0.90 | 3.77 | **1.80** | -0.39 |
| *ehu.es* | 6 | 4 | 1 | 1 | 2.13 | 1.93 | 0.45 | 1.89 | **1.60** | -1.44 |

*Menéame*

Regarding the Menéame service, the results are higher than expected, especially for the first snapshot in March, but after then there is a significant drop in the data retrieved (table 7).

Table 7. Inlinks and Rv for Meneame

| DOMAIN | LINKS | | | | RELATIVE REPRESENTATION | | | | Rv | r (%) |
|---|---|---|---|---|---|---|---|---|---|---|
| | MAR | JUN | SEP | DEC | MAR | JUN | SEP | DEC | | |
| *uib.es* | 826 | 586 | 449 | 544 | 15.27 | 31.10 | 28.69 | 31.52 | **26.64** | -0.40 |
| *upc.es* | 306 | 87 | 74 | 92 | 5.66 | 4.62 | 4.73 | 5.33 | **5.08** | -1.04 |
| *ie.edu* | 257 | 94 | 62 | 42 | 4.75 | 4.99 | 3.96 | 2.43 | **4.03** | -1.46 |
| *us.es* | 242 | 84 | 67 | 47 | 4.47 | 4.46 | 4.28 | 2.72 | **3.98** | -1.34 |
| *ugr.es* | 218 | 61 | 77 | 55 | 4.03 | 3.24 | 4.92 | 3.19 | **3.84** | -1.17 |
| *ucm.es* | 229 | 64 | 56 | 66 | 4.23 | 3.40 | 3.58 | 3.82 | **3.76** | -1.07 |
| *upm.es* | 249 | 52 | 62 | 55 | 4.60 | 2.76 | 3.96 | 3.19 | **3.63** | -1.26 |
| *uv.es* | 182 | 51 | 41 | 33 | 3.36 | 2.71 | 2.62 | 1.91 | **2.65** | -1.39 |
| *uvigo.es* | 129 | 26 | 27 | 77 | 2.38 | 1.38 | 1.73 | 4.46 | **2.49** | -0.48 |
| *deusto.es* | 47 | 25 | 39 | 67 | 0.87 | 1.33 | 2.49 | 3.88 | **2.14** | 0.37 |

First, and very prominently, comes "uib.es", followed by "upc.es". The data for the 10 domains with the greatest Rv factor are shown below. It is important to note that UIB is the university where the creator of this platform (Ricardo Galli) works, as this explains the impressive performance of this institution on Menéame, and probably on Digg too.

**c) News services**

*SINC platform*

The results from SINC Platform are shown in table 8.

Table 8. Inlinks and Rv for SINC platform

| DOMAIN | LINKS | | | | RELATIVE REPRESENTATION | | | | Rv | r (%) |
|---|---|---|---|---|---|---|---|---|---|---|
| | MAR | JUN | SEP | DEC | MAR | JUN | SEP | DEC | | |
| *upm.es* | 289 | 120 | 251 | 234 | 13.35 | 11.85 | 13.07 | 12.07 | **12.58** | -0.21 |
| *uc3m.es* | 200 | 59 | 210 | 261 | 9.24 | 5.82 | 10.93 | 13.46 | **9.86** | 0.28 |
| *uam.es* | 223 | 64 | 157 | 162 | 10.30 | 6.32 | 8.17 | 8.35 | **8.29** | -0.31 |
| *ugr.es* | 185 | 70 | 175 | 132 | 8.55 | 6.91 | 9.11 | 6.81 | **7.84** | -0.32 |
| *ub.edu* | 86 | 74 | 154 | 83 | 3.97 | 7.31 | 8.02 | 4.28 | **5.89** | -0.04 |
| *ucm.es* | 128 | 35 | 78 | 99 | 5.91 | 3.46 | 4.06 | 5.11 | **4.63** | -0.25 |
| *upf.edu* | 82 | 33 | 68 | 150 | 3.79 | 3.26 | 3.54 | 7.74 | **4.58** | 0.65 |
| *upc.edu* | 100 | 36 | 72 | 56 | 4.62 | 3.55 | 3.75 | 2.89 | **3.70** | -0.54 |
| *uniovi.es* | 50 | 28 | 47 | 45 | 2.31 | 2.76 | 2.45 | 2.32 | **2.46** | -0.10 |
| *unav.es* | 54 | 27 | 46 | 41 | 2.49 | 2.67 | 2.39 | 2.11 | **2.42** | -0.27 |

Up to 62 URLs (43.97%) receive no links from the platform SINC, which represents a high figure. However, for universities that do receive links, their number is higher than expected.

There is a clear dominance of universities based in Madrid, occupying the top 3 positions (UPM, UC3M and UAM), and sixth (UCM).

Furthermore, a very substantial data decrease is detected in June, which is recovered in subsequent measurements. This could be due to lower activity in the summer months.

**d) Resource sharing systems**

As regards platforms for sharing resources, the following services are studied: Delicious, Flickr, Slideshare, and Youtube.

*Delicious*

The results provided by Delicious are also higher than expected (table 9), obtaining -in the last available sample- up to 38 URLs (26.95%) with more than one thousand bookmarks pointing to them. The good performance of the Catalan universities should also be noted, with 4 domains in the top ten, as well as Madrid (UCM and UPM, the first and third places respectively).

**Table 9. Bookmarks and Rv for Delicious**

| DOMAIN | LINKS | | | | RELATIVE REPRESENTATION | | | | Rv | r (%) |
|---|---|---|---|---|---|---|---|---|---|---|
| | MAR | JUN | SEP | DEC | MAR | JUN | SEP | DEC | | |
| *ucm.es* | 4,609 | 5,078 | 5,298 | 5,796 | 4.79 | 4.86 | 5.07 | 4.95 | **4.92** | -0.22 |
| *uoc.edu* | 4,037 | 4,531 | 4,691 | 5,413 | 4.19 | 4.33 | 4.48 | 4.63 | **4.41** | -0.28 |
| *upm.es* | 4,227 | 4,499 | 4,652 | 4,878 | 4.39 | 4.30 | 4.45 | 4.17 | **4.33** | -0.14 |
| *us.es* | 4,002 | 4,318 | 4,459 | 4,844 | 4.16 | 4.13 | 4.26 | 4.14 | **4.17** | -0.19 |
| *ub.edu* | 3,161 | 3,734 | 3,849 | 4,060 | 3.28 | 3.57 | 3.68 | 3.47 | **3.50** | -0.24 |
| *upf.edu* | 3,459 | 3,624 | 3,693 | 3,983 | 3.59 | 3.47 | 3.53 | 3.40 | **3.50** | -0.14 |
| *ugr.es* | 3,250 | 3,459 | 3,548 | 3,782 | 3.37 | 3.31 | 3.39 | 3.23 | **3.33** | -0.15 |
| *upv.es* | 2,994 | 3,206 | 3,313 | 3,599 | 3.11 | 3.07 | 3.17 | 3.08 | **3.10** | -0.18 |
| *unirioja.es* | 2,672 | 3,184 | 3,315 | 3,764 | 2.77 | 3.05 | 3.17 | 3.22 | **3.05** | -0.33 |
| *uab.es* | 2,976 | 3,135 | 3,228 | 3,441 | 3.09 | 3.00 | 3.09 | 2.94 | **3.03** | -0.14 |

The results have the distinction of being cumulative, because the service is based on recording users' favorite URLs, so that the system accumulates the existing URLs and adds the new records. The domains "ucm.es", "uoc.edu" and "unirioja.es" are those which registered a positive upward trend. Although the system does not record a decrease in data at any time, in September, a number of URLs with 0 results are detected ("uniovi.es", "deusto.es", "mondragon.edu", "upcomillas.es", "upo.es", and "ufv.es"), due to a specific problem in Delicious at the time of data collection, which disappears in the next snapshot.

*Flickr*

The data obtained for Flickr is shown in table 10, displaying very low values. No URL receives more than 100 links from this service, although a gradual growth over the months (from 330 links in March up to 557 in December) is detected. It also highlights the high values for "unileon.es" and "unia.es", web domains with lower performance in other indicators.

Table 10. Inlinks and Rv for Flickr

| DOMAIN | LINKS | | | | RELATIVE REPRESENTATION | | | | Rv | r (%) |
|---|---|---|---|---|---|---|---|---|---|---|
| | MAR | JUN | SEP | DEC | MAR | JUN | SEP | DEC | | |
| us.es | 34 | 39 | 62 | 86 | 10.30 | 10.29 | 12.35 | 15.44 | **12.10** | 1.04 |
| upf.edu | 39 | 32 | 48 | 47 | 11.82 | 8.44 | 9.56 | 8.44 | **9.57** | 0.19 |
| ugr.es | 20 | 19 | 27 | 30 | 6.06 | 5.01 | 5.38 | 5.39 | **5.46** | 0.43 |
| unileon.es | 14 | 20 | 24 | 28 | 4.24 | 5.28 | 4.78 | 5.03 | **4.83** | 0.76 |
| upm.es | 10 | 22 | 24 | 24 | 3.03 | 5.80 | 4.78 | 4.31 | **4.48** | 0.98 |
| uv.es | 21 | 12 | 17 | 14 | 6.36 | 3.17 | 3.39 | 2.51 | **3.86** | -0.39 |
| uoc.edu | 13 | 11 | 20 | 22 | 3.94 | 2.90 | 3.98 | 3.95 | **3.69** | 0.56 |
| um.es | 15 | 14 | 16 | 16 | 4.55 | 3.69 | 3.19 | 2.87 | **3.57** | 0.07 |
| unia.es | 6 | 13 | 15 | 22 | 1.82 | 3.43 | 2.99 | 3.95 | **3.05** | 1.54 |
| ub.es | 9 | 9 | 15 | 15 | 2.73 | 2.37 | 2.99 | 2.69 | **2.70** | 0.54 |

*Slideshare*

The results are also very discrete on Slideshare (table 11), where the 73 links to "uoc.edu" in December (adding 14 more that "uoc.es" receives) should be mentioned. The results, as well as on Flickr, grow during all the months studied. Also noteworthy are the relatively high positions achieved by the private universities. In addition to UOC, UDIMA and UD are near the top and, slightly behind, IE University in fourteenth position.

Table 11. Inlinks and Rv for Slideshare

| DOMAIN | LINKS | | | | RELATIVE REPRESENTATION | | | | Rv | r (%) |
|---|---|---|---|---|---|---|---|---|---|---|
| | MAR | JUN | SEP | DEC | MAR | JUN | SEP | DEC | | |
| uoc.edu | 55 | 66 | 75 | 73 | 16.13 | 14.38 | 13.51 | 12.67 | **14.17** | 0.29 |
| uvigo.es | 12 | 17 | 21 | 20 | 3.52 | 3.70 | 3.78 | 3.47 | **3.62** | 0.54 |
| ua.es | 12 | 15 | 18 | 20 | 3.52 | 3.27 | 3.24 | 3.47 | **3.38** | 0.54 |
| ub.edu | 13 | 16 | 18 | 16 | 3.81 | 3.49 | 3.24 | 2.78 | **3.33** | 0.21 |
| udima.es | 10 | 14 | 17 | 18 | 2.93 | 3.05 | 3.06 | 3.13 | **3.04** | 0.63 |
| us.es | 11 | 13 | 14 | 16 | 3.23 | 2.83 | 2.52 | 2.78 | **2.84** | 0.39 |
| upv.es | 10 | 12 | 13 | 19 | 2.93 | 2.61 | 2.34 | 3.30 | **2.80** | 0.70 |
| deusto.es | 10 | 11 | 15 | 16 | 2.93 | 2.40 | 2.70 | 2.78 | **2.70** | 0.50 |
| uoc.es | 7 | 12 | 15 | 14 | 2.05 | 2.61 | 2.70 | 2.43 | **2.45** | 0.76 |
| uma.es | 6 | 13 | 14 | 14 | 1.76 | 2.83 | 2.52 | 2.43 | **2.39** | 0.94 |

*Youtube*

The results obtained from Youtube are somewhat higher than those corresponding to other file sharing services discussed above. Despite this, only 2 domains get over 100 links from Youtube in the last data sample. As in the above services, the trend is upward, from 868 links in March to 1,303 in December 2010 (already shown in table 2).

The main technical universities (UPC, UPM and UPV) achieve good results (second, fifth and sixth places respectively). In addition, some private universities, such as UNAV, and IE, find their way into the top ten (table 12).

Table 12. Inlinks and Rv for Youtube

| DOMAINS | LINKS | | | | RELATIVE REPRESENTATION | | | | Rv | r (%) |
|---|---|---|---|---|---|---|---|---|---|---|
| | MAR | JUN | SEP | DEC | MAR | JUN | SEP | DEC | | |
| *upf.edu* | 149 | 113 | 140 | 169 | 17.17 | 14.95 | 13.42 | 12.97 | **14.63** | 0.13 |
| *upc.edu* | 37 | 56 | 100 | 135 | 4.26 | 7.41 | 9.59 | 10.36 | **7.90** | 1.53 |
| *unav.es* | 41 | 42 | 66 | 64 | 4.72 | 5.56 | 6.33 | 4.91 | **5.38** | 0.47 |
| *ie.edu* | 44 | 39 | 55 | 71 | 5.07 | 5.16 | 5.27 | 5.45 | **5.24** | 0.51 |
| *upm.es* | 36 | 41 | 56 | 68 | 4.15 | 5.42 | 5.37 | 5.22 | **5.04** | 0.69 |
| *upv.es* | 44 | 44 | 51 | 55 | 5.07 | 5.82 | 4.89 | 4.22 | **5.00** | 0.23 |
| *us.es* | 43 | 20 | 20 | 23 | 4.95 | 2.65 | 1.92 | 1.77 | **2.82** | -0.58 |
| *uoc.edu* | 26 | 15 | 22 | 29 | 3.00 | 1.98 | 2.11 | 2.23 | **2.33** | 0.11 |
| *uv.es* | 21 | 15 | 19 | 27 | 2.42 | 1.98 | 1.82 | 2.07 | **2.07** | 0.26 |
| *ull.es* | 20 | 14 | 22 | 26 | 2.30 | 1.85 | 2.11 | 2.00 | **2.07** | 0.27 |

**e) Reference information**

*Wikipedia*

The social encyclopedia Wikipedia is the service that provides the most external links to the set of academic Spanish domains in 2010. Table 13 includes, as in all other services analyzed thus far, the total number of links received, the normalized values and the Rv factor. The table also reflects the preponderance of 3 domains: "ucm.es" (Rv=15.32), "us.es" (Rv=13.27), and "unirioja.es" (Rv=11.97), corresponding to the three largest domains according to count size (Annex 2).

The fourth domain (corresponding to "uv.es", with 2,270 external links in December 2010), is already far behind these top three domains (if compared to 6,140 links to "unirioja.es").

Table 13. Inlinks and Rv for Wikipedia

| DOMAINS | LINKS | | | | RELATIVE REPRESENTATION | | | | Rv | r (%) |
|---|---|---|---|---|---|---|---|---|---|---|
| | MAR | JUN | SEP | DEC | MAR | JUN | SEP | DEC | | |
| *ucm.es* | 5,960 | 6,620 | 7,520 | 7,370 | 15.93 | 13.36 | 16.15 | 15.85 | **15.32** | 0.22 |
| *us.es* | 5,610 | 5,620 | 6,440 | 6,010 | 15.00 | 11.34 | 13.83 | 12.92 | **13.27** | 0.07 |
| *unirioja.es* | 4,390 | 4,871 | 6,100 | 6,140 | 11.74 | 9.83 | 13.10 | 13.20 | **11.97** | 0.35 |
| *uv.es* | 1,820 | 2,270 | 2,270 | 2,270 | 4.87 | 4.58 | 4.87 | 4.88 | **4.80** | 0.23 |
| *ub.es* | 1,240 | 1,690 | 1,630 | 1,600 | 3.32 | 3.41 | 3.50 | 3.44 | **3.42** | 0.26 |
| *ehu.es* | 1,150 | 1,660 | 1,490 | 1,470 | 3.07 | 3.35 | 3.20 | 3.16 | **3.20** | 0.25 |
| *uoc.edu* | 1,100 | 1,610 | 1,310 | 1,300 | 2.94 | 3.25 | 2.81 | 2.80 | **2.95** | 0.17 |

| | | | | | | | | | |
|---|---|---|---|---|---|---|---|---|---|
| *ua.es* | 861 | 1,320 | 1,100 | 1,110 | 2.30 | 2.66 | 2.36 | 2.39 | **2.43** | 0.26 |
| *uib.es* | 836 | 1,350 | 1,060 | 1,080 | 2.24 | 2.72 | 2.28 | 2.32 | **2.39** | 0.26 |
| *unizar.es* | 825 | 1,230 | 1,100 | 1,110 | 2.21 | 2.48 | 2.36 | 2.39 | **2.36** | 0.31 |

**f) Blogging and microblogging**

This section focuses on the analysis of Technorati (blog search engine) and Twitter (microblogging tool).

*Technorati*

Table 14 shows the results for Technorati, which confirms the lack of influence of this service in providing web links to the Spanish academic space.

The most linked domain is "ie.edu", but this suffers an important loss of links over time, from 52 in June to only 9 links in December. In this last snapshot only 15 domains receive at least one link, the remaining URLs do not collect any.

**Table 14. Inlinks and Rv for Technorati**

| DOMAINS | LINKS | | | | RELATIVE REPRESENTATION | | | | Rv | r (%) |
|---|---|---|---|---|---|---|---|---|---|---|
| | MAR | JUN | SEP | DEC | MAR | JUN | SEP | DEC | | |
| *ie.edu* | 52 | 6 | 6 | 9 | 61.18 | 24.00 | 20.69 | 25.00 | **32.72** | -1.42 |
| *upv.es* | 4 | 5 | 4 | 3 | 4.71 | 20.00 | 13.79 | 8.33 | **11.71** | -0.28 |
| *us.es* | 10 | 4 | 3 | 3 | 11.76 | 16.00 | 10.34 | 8.33 | **11.61** | -1.04 |
| *mondragon.edu* | 2 | 4 | 2 | 2 | 2.35 | 16.00 | 6.90 | 5.56 | **7.70** | 0.00 |
| *uam.es* | 0 | 0 | 4 | 4 | 0.00 | 0.00 | 13.79 | 11.11 | **6.23** | 0 |
| *uc3m.es* | 0 | 0 | 1 | 6 | 0.00 | 0.00 | 3.45 | 16.67 | **5.03** | 0 |
| *unizar.es* | 3 | 1 | 1 | 1 | 3.53 | 4.00 | 3.45 | 2.78 | **3.44** | -0.96 |
| *uib.es* | 2 | 1 | 1 | 1 | 2.35 | 4.00 | 3.45 | 2.78 | **3.14** | -0.64 |
| *um.es* | 4 | 1 | 0 | 1 | 4.71 | 4.00 | 0.00 | 2.78 | **2.87** | -1.17 |
| *udl.cat* | 0 | 0 | 3 | 0 | 0.00 | 0.00 | 10.34 | 0.00 | **2.59** | 0 |

*Twitter*

Finally, table 15 shows the results obtained through the microblogging service Twitter.

The results are very discrete, but the nature of this service should be taken into account (140 characters, which hinders the introduction of URLs) as well as obsolescence and the volatility of tweets. However, an increasing trend is detected, reflected in the increase of the cumulative total number of links, from 311 in March up to 637 in December.

Lastly, the performance of three private universities (UOC, UNAV and IE) at the top should be noted, and that of 3 polytechnics (UPV, and UPM, and UPC).

Table 15. Inlinks and Rv for Twitter

| DOMAIN | LINKS | | | | RELATIVE REPRESENTATION | | | | Rv | r (%) |
|---|---|---|---|---|---|---|---|---|---|---|
| | MAR | JUN | SEP | DEC | MAR | JUN | SEP | DEC | | |
| *uoc.edu* | 18 | 28 | 30 | 36 | 5.79 | 6.53 | 5.79 | 5.65 | **5.94** | 0.76 |
| *us.es* | 15 | 23 | 30 | 34 | 4.82 | 5.36 | 5.79 | 5.34 | **5.33** | 0.91 |
| *unav.es* | 16 | 21 | 24 | 25 | 5.14 | 4.90 | 4.63 | 3.92 | **4.65** | 0.47 |
| *upv.es* | 13 | 19 | 25 | 32 | 4.18 | 4.43 | 4.83 | 5.02 | **4.61** | 1.01 |
| *ie.edu* | 15 | 21 | 23 | 23 | 4.82 | 4.90 | 4.44 | 3.61 | **4.44** | 0.45 |
| *uv.es* | 17 | 14 | 21 | 24 | 5.47 | 3.26 | 4.05 | 3.77 | **4.14** | 0.36 |
| *ugr.es* | 7 | 13 | 12 | 40 | 2.25 | 3.03 | 2.32 | 6.28 | **3.47** | 2.18 |
| *upm.es* | 11 | 15 | 19 | 19 | 3.54 | 3.50 | 3.67 | 2.98 | **3.42** | 0.59 |
| *uji.es* | 12 | 14 | 15 | 19 | 3.86 | 3.26 | 2.90 | 2.98 | **3.25** | 0.49 |
| *upc.edu* | 8 | 12 | 17 | 21 | 2.57 | 2.80 | 3.28 | 3.30 | **2.99** | 1.09 |

### 4.2. Correlation analysis

Annex 4 (containing tables 19 to 22) shows the Spearman correlation for each platform analyzed, for each of the four samples considered. External inlink correlation with each of these platforms is additionally added.

The correlation values obtained are unexpectedly high (given the low percentage that selective links represent with respect to total links). Only for Digg and Technorati are the results not meaningful. These are the platforms which retrieved the fewest raw links. This serves to illustrate the fact that the services with the highest number of links to the Spanish system (as displayed in table 2) are those that achieve better correlations.

Among the platforms which provide the most links (Wikipedia, Delicious, Linkedin, and Academia), the correlation is very high. In December, between Delicious and Wikipedia the correlation is 0.961 (the highest value obtained between platforms). Delicious also correlates accurately with Linkedin (0.921), and Academia (0.805).

Furthermore, Annex 5 (table 23) shows the evolution, over time, of the correlation between the external links and each of the platforms studied. Results confirm the high values obtained in all four samples. Only Digg and Technorati do not achieve representative correlation. In order to place the focus on the URLs with higher performance, table 24 (Annex 5) represents the same correlation, but taking into account only the URLs with more than 50,000 external inlinks (strong URLs as regards web visibility).

The results obtained show some differences in performance. The high correlation both of Delicious and Wikipedia is maintained, but values for all other platforms are reduced significantly. For example, looking at data for December, the overall correlation between external links and links from Youtube is 0.854, but if we consider only the top URLs, this value is reduced to 0.491. A similar effect is detected for Slideshare and Sinc.

The four platforms with the best correlation in December, taking only into account URLs with more than 50,000 links, are Delicious (0.870), Wikipedia (0.832), Linkedin (0.693), and Facebook (0.639). For all these platforms, the regression analysis against external inlinks is provided in figure 3.

**Figure 3. Correlation between external links and selective sites (Delicious, Wikipedia, Linkedin, and Facebook)**

As can be observed, the low area of each distribution shows a significant dispersion of data, and in the upper zone (URLs with more external links, and more selective links), the correlation is almost perfect, especially in Delicious and Facebook.

## 5. Discussion

The percentage of links from social sites to Spanish universities is very small. In December, only 1.72% of total incoming external links to the 76 Spanish universities came from social platforms. Those that do stand out -in relative terms- are the links received from Delicious (particularly in the case of UCM, with 5,796 links in December), and Wikipedia (where UCM is also the most representative domain, with 7,370 inlinks in December).

Other platforms showing upward trends are Linkedin (which grows from 2,089 links in March to 2,690 in December) and, to a lesser extent, Flickr (557 accumulated links in December), Youtube (1,043) and Twitter (518), although they grow very slowly and with very small numbers of links. Moreover, a negative trend is detected in Facebook, whereby after a large increase in links between March and June (from 2,777 accumulated links to 5,037), they subsequently decrease (4,253 in December).

The social news manager Menéame presents one of the most negative trends, from the 5,411 accumulated links in March to 1,726 in December, while its counterpart Digg has very little coverage of Spanish universities (providing only 53 links in December). On the other hand, the news platform SINC shows a higher number of links than expected, maintaining around 2,000 links for all the snapshots, except June, which produces a significant drop that is later recovered.

As regards the relative position of universities on different sites, on Wikipedia the 3 domains with the most links from this service, are "ucm.es" ($Rv = 15.32$), "us.es" ($Rv = 13.27$) and "unirioja.es" ($Rv = 11.97$), precisely the 3 domains with largest web spaces. In fact, Delicious (the platform with the second largest number of links to universities), exhibits similar behavior: "ucm.es" is the most representative domain ($Rv = 4.92$), "us.es" the fourth ($Rv = 4.17$) and "unirioja.es" further back, the ninth ($Rv = 3.05$).

Moreover, "unirioja.es" scores highly on Academia.edu ($Rv = 10.04$); "ub.edu" on Facebook ($Rv = 12.64$); "uib.es" on Digg ($Rv = 35.04$) and Menéame ($Rv = 26.64$); "ie.edu" on Linkedin ($Rv = 7.22$) and Technorati ($Rv = 32.72$); "uoc.edu" on Twitter ($Rv =

5.94) and Slideshare (Rv = 14.17); "us.es" on Flickr; and "upf.edu" on Youtube (Rv = 14.63).

The correlation results show very high values between total external links and each platform. Moreover, between the platforms, the correlations are also extremely good. Only Digg (low representation of Spanish Universities) and Technorati (decrease of usage of this platform) achieve non-representative correlation. Despite this, if only strong web domains are considered (those with more than 50,000 results per sample), the results change significantly, and only Delicious and Wikipedia maintain good correlation values.

Nonetheless, these results should be contextualized by size (page count), due to the direct relation between size and visibility. In this sense, the larger domains (such as "ucm.es", "us.es") also show high Rv values both in Wikipedia and Delicious, and an elevated number of total inlinks. This performance is generally detected for all domains with a high page count.

In any case, the page count of the larger domains is explained by the existence of special services, such as Compludoc ("ucm.es"), and Dialnet ("unirioja.es"), an effect previously detected by Orduña-Malea et al. (2010), and whose direct influence in the generation of total external and selective links should be further analyzed in future studies.

In broad terms, old, well-established, general universities (those with high page counts and external inlinks, such as UB, UCM, UGR, UV, US, etc.) perform very well in all indicators, while private universities stand out on some specific websites, indicating a focused effort in the usage of these services, such as IE on Linkedin (as a business strategy) or UOC on Slideshare (as a teaching service).

The high correlation obtained by Wikipedia could reinforce this possible advantage of older (more time creating content) and bigger (more faculties, departments, colleges, and infrastructure) universities, since these institutions are statistically more likely to have more Wikipedia entries, and to be more frequently mentioned and/or linked.

## 6. Conclusions

The main conclusions are set out below:

- The percentage of links from social sites is very small (in December only 1.72%). In any case, this percentage presents a positive trend (is expected to grow), and it should also be considered that the existence of links in some of these services can be unusual for their users.

- Some private universities (UOC and IE University) achieve good results (receive a large amount of links from some social platforms), implying the existence of specific positioning policies that are providing web visibility for these universities.

- Despite the low percentage of links, the correlation between links from each platform and total external links is very high, but if we consider only the domains with the most external links, only Delicious and Wikipedia provide good level correlations, and can be used as good substitutes.

In any case, all results obtained should be contextualized, taking the following issues into account:

- The purpose of the platforms analyzed is not to provide links to universities; indeed in some cases the inclusion of links is either unusual or unfriendly.

- The creation of channels on some platforms by universities might influence the amount of links that are received from these platforms.

Moreover, the results have been obtained via the analysis of a specific set of universities (the Spanish system) and platforms. This analysis should be applied to other university systems and complemented by the addition of new social platforms to corroborate the detected patterns.

As the services used to gather the web data have disappeared, further research is needed in order to check the results obtained, using other alternative indicators, such as URL mentions.

## 7. Notes

1. http://www.wiserweb.org. Accessed 4 June 2012.
2. http://www.webindicators.org. Accessed 4 June 2012.
3. http://www.eicstes.org. Accessed 4 June 2012.
4. http://www.educacion.es/educacion/universidades/educacion-superior-universitaria/que-estudiar-donde/universidades-espanolas.html. Accessed 4 June 2012.
5. http://www.crue.org. Accessed 4 June 2012.
6**.** http://altmetrics.org. Accessed 4 June 2012.